\documentclass[preprint,nofootinbib,aps]{revtex4}
\usepackage{graphicx}
\begin{document}

\title{The Phantom Bounce: A New Oscillating Cosmology}

\author{Matthew G. Brown, Katherine Freese}
\affiliation{Michigan Center for Theoretical Physics, Dept. of Physics,\\
   University of Michigan, Ann Arbor, MI 48109. Email: {\tt brownmg@umich.edu,
     ktfreese@umich.edu}} 
\author{William H. Kinney}
\affiliation{Dept.\ of Physics, Univerity at Buffalo, SUNY, Buffalo,
   NY 14260. Email: {\tt whkinney@buffalo.edu}}
 
\begin{abstract} 
   An oscillating universe cycles through a series of expansions and
   contractions.  We propose a model in which ``phantom'' energy with $p <  - \rho$
   grows rapidly and dominates the late-time expanding phase.  The
   universe's energy density is so large that the effects of quantum
   gravity are important at both the beginning and the
   end of each expansion (or contraction).  The bounce can be caused by
   high energy modifications to the Friedmann equation, which make the 
   cosmology nonsingular.  The classic 
   black hole overproduction of oscillating universes is resolved due 
   to their destruction by the phantom energy.

\end{abstract}

\maketitle  

\section{Introduction}

In this Letter, we consider a scenario in which the universe
oscillates through a series of expansions and contractions.  After it
finishes its current expanding phase, the universe reaches a state of
maximum expansion which we will call ``turnaround'', and then begins
to recollapse.  Once it reaches its smallest extent at the ``bounce'',
it will once again begin to expand. This scenario is distinguished
from other proposed cyclic universe scenarios \cite{tolman,steinturok}
in that cosmological acceleration due to ``phantom'' energy ({\it
   i.e.}, dark energy with a supernegative equation of state, $p < -
\rho$) \cite{phantom} plays a crucial role.  In addition, our work differs from
recent proposals in that our model takes place in 3 space and 1 time
dimension (though the proposed mechanism for the bounce arises from
braneworld scenarios).

The idea of an oscillating universe was first proposed in the 1930's
by Tolman.  Over the subsequent decades, two problems stymied the
success of oscillating models. First, the formation of large scale
structure and of black holes during the expanding phase leads to
problems during the contracting phase \cite{dickepeebles}.  The black
holes, which cannot disappear due to Hawking area theorems,
grow ever larger during subsequent cycles. Eventually, they occupy the entire
horizon volume during the contracting phase so that calculations break
down.  (Only the smallest black holes can evaporate via Hawking
radiation.)  The second unsolved problem of oscillating models was the
lack of a mechanism for the bounce and turnaround.
The turnaround at the end of the
expanding phase might be explained by invoking a closed universe, but
the recent evidence for cosmological acceleration removes that
possibility. For the observationally favored density of ``dark
energy'', even a closed universe will expand forever. Thus, cyclic
cosmologies appeared to conflict with observations.

Our scenario resolves these problems.  Our resolution to the black
hole overproduction problem is provided by a ``phantom'' component to
the universe, which destroys all structures towards the end of the
universe's expanding stage.  Phantom energy, a proposed explanation
for the acceleration of the universe, is characterized by a component
$Q$ with equation of state
\begin{equation}
w_Q = p_Q/\rho_Q < -1 .
\end{equation}
Since the sum of the pressure and energy density is negative, the
dominant energy bound of general relativity is violated; yet recent
work explores such models nevertheless. Phantom energy can dominate
the universe today and drive the current acceleration. Then it becomes
ever more dominant as the universe expands.  With such an unusual
equation of state, the Hawking area theorems fail, and black holes can
disappear \cite{davies}.  In ``big rip'' scenarios \cite{ckw}, the
rapidly accelerating expansion due to this growing phantom component
tears apart all bound objects including black holes.  (We speculate
about remnants of these black holes below.)

The phantom energy density becomes infinite in finite time
\cite{caldwell,ckw}.  The energy density of any field described by
equation of state $w_Q$ depends on the scale factor $a$ as
\begin{equation}
\label{eq:grow}
\rho_Q \sim a^{-3(1+w_Q)} .
\end{equation}
Hence, for $w_Q<-1$, $\rho_Q$ grows as the universe expands.  Of
course, we expect that an epoch of quantum gravity sets in before the
energy density becomes infinite.  We therefore arrive at the peculiar
notion that quantum gravity governs the behavior of the universe both
at the beginning and at the end of the expanding universe (i.e., at
the smallest and largest values of the scale factor). Here we consider
an example of the role that high energy density physics may play on
both ends of the lifetime of an expanding universe: we consider the
idea that large energy densities may cause the universe to bounce when
it is small, and to turn around when it is large. The idea is
economical in that it is the {\em same physics} which operates at both
bounce and turnaround.

In this Letter we use modifications to the Friedmann equations to
provide a mechanism for the bounce and the turnaround that are
responsible for the alternating expansion and contraction of the
universe. In particular, we focus on ``braneworld'' scenarios in
which our observable universe is a three-dimensional surface situated 
in extra dimensions. Several scenarios for implementing a bounce have
been proposed in the literature \cite{shtanov,branebounce}. As an example,
we focus on the modification to the Randall-Sundrum \cite{RSI}
scenario proposed by Shtanov and Sahni \cite{shtanov}, 
which involves a negative brane tension and a timelike 
extra dimension leading to a modified Friedmann equation. 
Another example is the quantum bounce in loop quantum gravity
\cite{singh}.
Once the energy density of the universe reaches a critical value, 
cosmological evolution changes direction: if it has been expanding, 
it turns around and begins to recontract. If it has been contracting, 
it bounces and begins to expand. 

We emphasize that the two components we propose here work together: we
use a modified Friedmann equation as a mechanism for a bounce and
turnaround, and we add a phantom component to the universe to destroy
black holes.  Due to the phantom component, the same high energy
behavior that produces a bounce at the end of the contracting phase
also produces a turnaround at the end of the expanding phase. In
addition, the bounce and turnaround are both nonsingular, unlike the
cyclic scenario proposed by Steinhardt and Turok \cite{steinturok},
which is complicated by a number of physical singularities related to
brane collisions near the bounce \cite{singularity}. This is currently
a very controversial topic.

\section{The Bouncing Cosmology} 

In an oscillating cosmology, what we observe to be ``The Big Bang'' really
is the universe emerging from a bounce.  The universe at this point
has its smallest extent (smallest scale factor $a$) and largest energy
density, somewhere near the Planck density.  The universe then
expands, its density decreases, and it goes through the classic
radiation dominated and matter dominated phases, with the usual
primordial nucleosynthesis, microwave background, and formation of
large structure. A period of inflation may or may not be necessary to
establish flatness and homogeneity.  At a redshift $z=O(1)$, the
universe starts to accelerate due to the existence of a vacuum
component or quintessence field $Q$. We take a ``phantom'' component
with $w_Q<-1$.  The energy density of this component grows rapidly as
the universe expands. Any structures produced during the expanding
phase, including galaxies and black holes, are torn apart by the
extremely rapid expansion provided by the phantom component.  Any
physics relevant at the high densities near the ``Big Bang'' again
becomes important at the high densities near the end of the expanding
phase. Modifications to the Friedmann equation  become important at 
high densities, and cause
the universe to turn around. The universe reaches a characteristic
maximum density 2$|\sigma|$ (which might be anywhere in the range from
TeV to $M_{p}$), and starts to contract.  As it contracts, at first
its energy density decreases (as the phantom component decreases in
importance), but then it again increases as matter and radiation
become dominant.  Eventually it reaches the high values at which the
modifications to Friedmann equations become important.  Once the
energy density again reaches the same characteristic scale
$2|\sigma|$, the universe stops contracting, bounces, and once again
expands.

In the standard cosmology, there is no way to avoid a singularity for
small radius or scale factor $a$.  In the context of extra dimensions,
however, one can have a bounce at finite $a$ so that singularities are
avoided.  A nonsingular bounce is obtained if the Friedmann equation
is modified by the addition of a new negative term on the right hand side:
\begin{equation}
\label{eq:hubble}
H^2 = {8 \pi \over 3 M_{p}^2} \left[\rho - f(\rho)\right], 
\end{equation}
where the function $f(\rho)$ is positive. For a contracting 
universe  to reverse and begin expanding again, we must have $\ddot a > 0$, 
which results in a condition on $f\left(\rho\right)$,
\begin{equation}
3 \left(1 + w\right)\rho f'\left(\rho\right) - 2 
f\left(\rho\right) - (1 + 3 w)\rho > 0.
\end{equation}
Similarly, for an expanding universe, 
$\ddot a$ must be negative for  the expansion to reverse.
A modified Friedmann equation of the form of Eq. (\ref{eq:hubble}) can
be motivated in the context of braneworld scenarios, where our
observable universe is a 3-dimensional surface embedded in extra
dimensions. Ref. \cite{cf} showed that Einstein's equations in higher
dimensions, together with Israel boundary conditions on our brane, can
give rise to an equation of the form of Eq. (\ref{eq:hubble}).
Different values of energy/momentum in the extra dimensions (the bulk)
can be responsible for different $f(\rho)$ in Eq.
(\ref{eq:hubble}).

In particular, we focus on ``braneworld'' motivated modifications to the
Friedmann equation, where the modification to the Friedmann equation for the
brane bound observer~\cite{shtanov,cf,braneworld} is
\begin{equation} 
H^2 =
{\Lambda_4 \over 3} + \left({8\pi \over 3 M_{p}^2} \right) \rho +
\epsilon \left({4\pi \over 3 M_5^3}\right)^2 \rho^2 + {C \over a^4}, 
\end{equation}
where the last term ($C$ is an integration constant) appears as
a form of ``dark radiation'' (that is constrained like ordinary  radiation),
and $\epsilon$ corresponds to the metric signature of the extra dimension
\cite{shtanov}.
We will also assume that the bulk cosmological 
constant is set so that the three-dimensional cosmological
constant $\Lambda_4$ is negligible\footnote{This fine-tuning is the
   usual cosmological constant problem, which is not addressed in this
   paper.}.  Hence the relevant correction to the Friedmann equation
  is the quadratic term, $f(\rho) =
\rho^2/2\left\vert\sigma\right\vert$. For $\epsilon < 0$, the Friedmann 
equation becomes
\begin{equation}
\label{eq:mod2}
H^2 = {8 \pi \over 3 M_{p}^2} \left[\rho - \frac{\rho^2}{2|\sigma|} \right].
\end{equation}
One way to obtain $\epsilon < 0$ corresponds to an 
extra {\em timelike} dimension: models with more than one time 
coordinate typically suffer from pathologies such as closed 
timelike curves and non-unitarity. We use the model in Ref. 
\cite{shtanov} to motivate the choice of sign in the Friedmann equation, 
but a more detailed treatment would need to address these other issues 
to form a fully consistent picture.  

Alternatively, 
in loop quantum gravity, there 
is a quantum bounce that takes place at Planck densities in lieu of the
singularity in the standard classical Friedmann equation \cite{singh}; 
if one couples
this quantum bounce with a phantom component as in this paper, one would
again obtain the same oscillating cosmology as discussed in this paper.

The expansion rate of the universe $H=0$ at 
$\rho_{bounce}=2|\sigma| $;
it is at this scale that the universe bounces and turns around. 
For this choice of $f\left(\rho\right)$,
\begin{equation}
3 \left(1 + w\right)\rho f'\left(\rho\right) - 2 f\left(\rho\right) - 
(1 + 3 w)\rho = 3 \left(1 + w\right) \rho,
\end{equation}
and the required condition on $\ddot a$ is 
satisfied at both bounce ($w > 0$) and turnaround ($w < -1$).
On one end of the cycle it goes from contracting to 
expanding (this bounce looks to us
like the Big Bang), and then at the other end of the cycle
it goes from expanding to contracting.
This behavior is illustrated in the Figure.
In models motivated by the Randall-Sundrum scenario, the most natural value 
of the brane tension is $\sigma = M_p$,
but we treat the problem generally for any value of $\sigma > {\rm TeV}$.

At scales above $\rho > \sigma$, the validity of
Eq. (\ref{eq:mod2}) breaks down in detail.  However, the approach to
$H=0$ and thus the existence of a bounce and turnaround remain
sensible.  In any case, we use this braneworld model merely as an example 
of a correction to the Friedmann equation. Other modifications to the
Friedmann equation might work as well, as long as there is the
requisite minus sign in the equation.

\begin{figure*}
\includegraphics[width=2.8in]{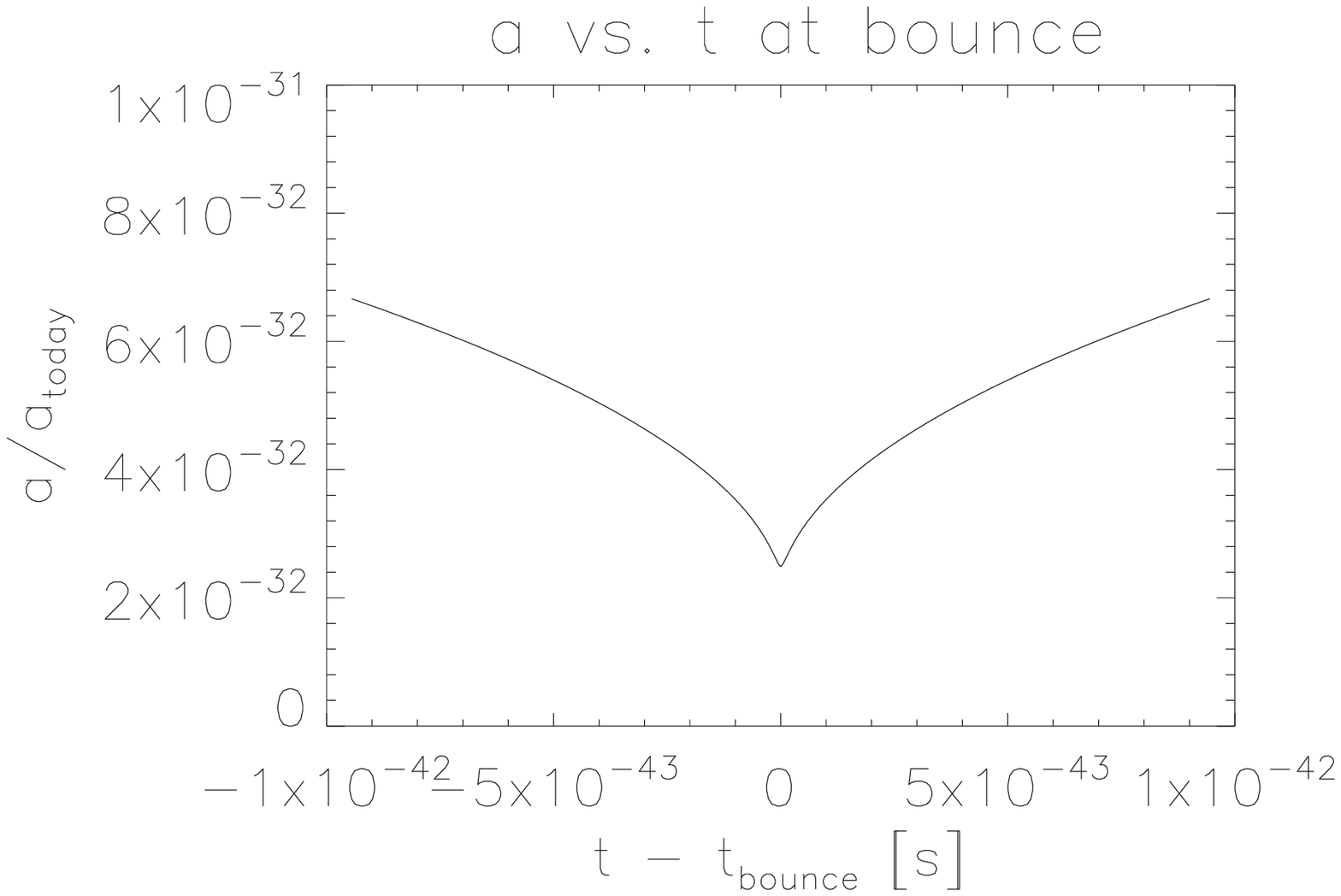}
\includegraphics[width=2.8in]{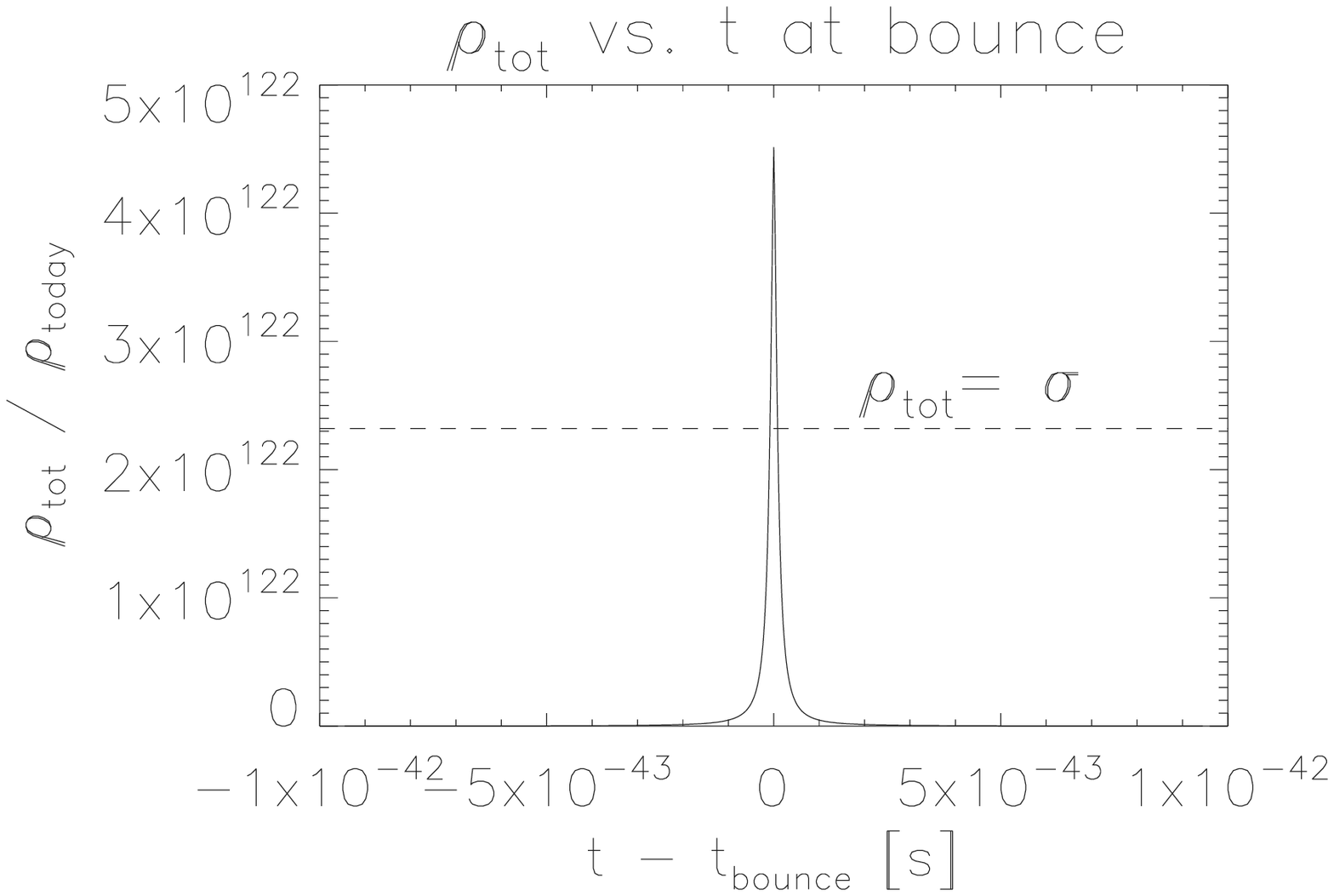} 
\includegraphics[width=2.8in]{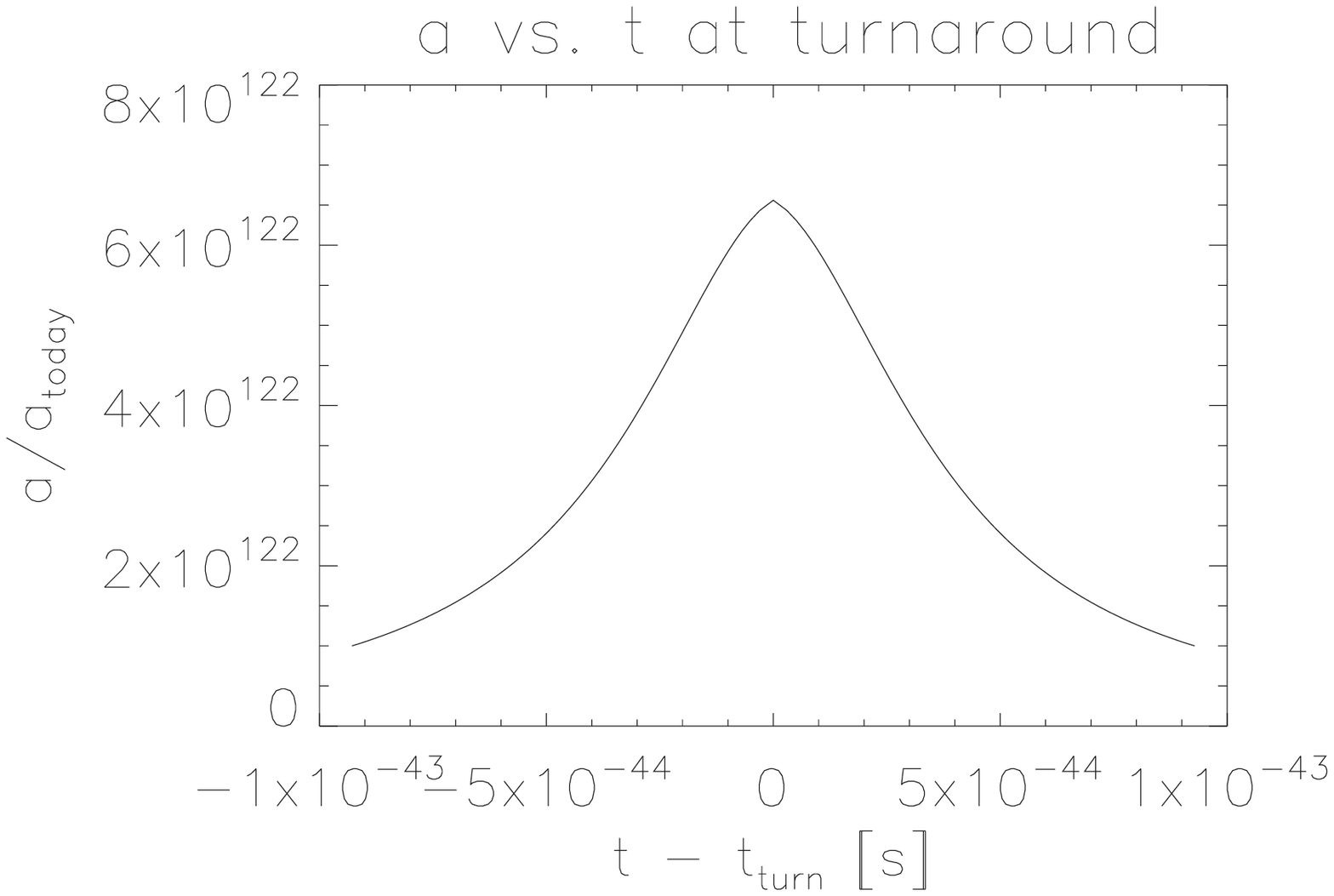}
\includegraphics[width=2.8in]{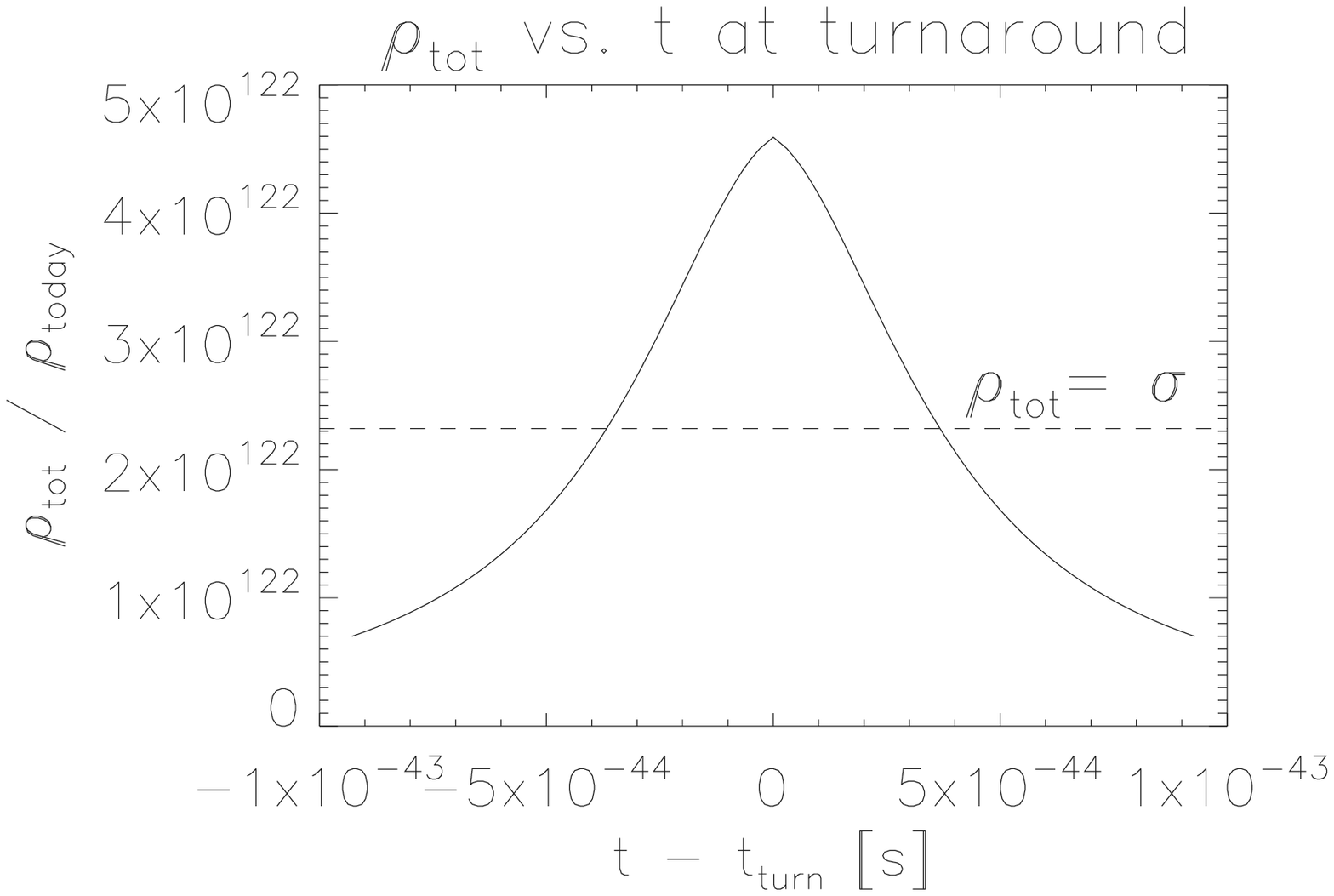}
\caption{Scale factor (left) and energy density (right) 
   at the bounce and turnaround, plotted as functions of time The
   dotted lines in the plot of the energy density show $\rho = \sigma$.
   We note that the energy density is large at the bounce due to
   radiation ($\rho_{\rm radn} \propto a^{-4}$) and is large at
   turnaround due to phantom energy $\rho_Q \propto a^{3|1+w_Q|}$.  The
   plots are presented for $w=-4/3$ to illustrate the basic behavior of
   the model (detailed numbers are irrelevant).}
\end{figure*}

\section{Destruction of Black Holes}

Black holes pose a serious problem in a standard oscillating universe.
However, the Hawking area theorems that guarantee the continued
existence of black holes have been constructed in special settings and
may not apply here; e.g., the same modifications to gravity that give
a bounce rather than a singularity in the cosmology may avoid
singularities in the black holes.  Indeed, when $w_Q < -1$, Davies
\cite{davies} has shown that the theorem does not hold. Recently, Caldwell,
{\it et al.}
\cite{ckw} described the dissolution of bound structures in the ``big
rip'' towards the end of a phantom dominated universe.  Any black
holes formed in an expanding phase of the universe are torn apart
before they can create problems during contraction.

When are the black holes destroyed?  We want to be certain that
they are torn apart before turnaround.  In
general relativity, the source for a gravitational potential is the
volume integral of $\rho + 3p$.  An object of
radius $R$ and mass $M$ is pulled apart when
\begin{equation}
-{4\pi \over 3}(\rho + 3p) R^3 \sim M .
\end{equation}
Writing $\rho + 3p = \rho(1+3w_Q)$ during phantom
domination and taking $R=2GM$ for the black hole,
we find that black holes are pulled apart when 
$-(4\pi / 3) \rho(1+3w_Q) 8 M^3/M_{p}^6 \sim M ,$
which happens when the energy density of the universe
has climbed to a value
\begin{equation}
\label{eq:rhobh}
\rho_{BH} \sim M_{p}^4 \left({M_{p}\over M}\right)^2 {3 \over 32\pi} {1 \over |1+3w_Q|} .
\end{equation}
More massive black holes are destroyed at lower values of $\rho$,
i.e. earlier.  It is the smallest black holes that get shredded last.

We must ensure that the black holes are destroyed before turnaround, so that 
$\rho_{BH} < \rho_{turn} = 2 |\sigma|$.  As an example, we can take $w_Q=-3$.
Then $10^6$ solar mass black holes, such as those at the centers of
galaxies, get pulled apart when $\rho \sim 10^{-90} M_{p}^4$, which easily
satisfies the above condition.  The most tricky case would be Planck
mass black holes, which either formed primordially or are relics of
larger black holes that Hawking radiated.  Even these should still be
disrupted.  From Eq. (\ref{eq:rhobh}) these will be shredded when $\rho
\sim 10^{-2}M_{p}$, before turnaround if the brane tension $|\sigma| =
M_{p}^4$. However, for GUT scale brane tension $|\sigma| = m_{GUT}^4$,
only black holes with $M \ge 10^{5} M_{p}$ are disrupted.  Fortunately
these black holes Hawking evaporate in a time $\tau \sim (25 \pi M^3
   / M_p^4)$ where $M$ is the black hole mass.  This occurs in only
$\sim 10^{-27}$ sec for a black hole with $M = 10^{5}M_{p}$.  We also
speculate that Planck mass remnant black holes that cannot disappear
(still containing the singularity) may be dark matter candidates.

\section{Discussion}

Our proposal contains the novel feature that both bounce 
and turnaround are produced by the same modification to the Friedmann 
equation. However, it does so at the price of including more than one 
speculative element: the modified Friedmann equation requires a
braneworld model to achieve, and the cosmology must be dominated by 
phantom energy. In many cases a phantom component
is difficult to implement from a fundamental 
standpoint without  severe pathologies such as an unstable vacuum 
(see, for example, Ref. \cite{Cline:2003gs}.) 
However, Parker and Raval \cite{parker} have investigated a
 cosmological model with zero cosmological constant, 
but containing the vacuum energy of a simple quantized 
free scalar field of low mass, and found that it has $w<-1$
without any pathologies.
Several additional areas also remain to be addressed. First, as the universe is
contracting, those modes of the density fluctuations that we usually
throw away as decaying (in an expanding universe) are instead growing.
Hence dangerous structures may form during the contracting phase.  At
the end of the contracting phase, there is no phantom energy to wipe
out whatever structure is formed. In this sense, the initial
conditions for structure formation in this picture are set either during the
phantom energy dominated epoch near turnaround or by the quantum generation of
fluctuations in the collapsing phase \cite{allen}. Black hole formation
could still kill the model.
  Second, it is not obvious that it is possible to
create a truly cyclic ({\it i.e.} perfectly periodic) cosmology within the context of the ``Phantom Bounce''
scenario. The reason for this is entropy production. We speculate that
it may be possible to create quasi-cyclic evolution by redshifting
entropy out of the horizon during the period of accelerating
expansion.  Even more speculatively, we note that the
special case of $w_Q = -7/3$, although disfavored by observation, 
possesses an intriguing duality between radiation 
($\rho_{rad} \propto a^{-4}$) and phantom energy 
($\rho_Q \propto a^4$). In this case, the behaviors of these components
exchange identity under a transformation $a \rightarrow 1/a$ \cite{duality}, effectively exchanging bounce for turnaround, a symmetry which might be exploited
to achieve truly cyclic evolution.

\section*{Acknowledgments}

We thank B. Burrington, J. Liu, R. McNees, and M. Trodden for 
conversations.  We acknowledge support from DOE as well as the Michigan Center 
for Theoretical Physics (MCTP) via the Univ. of Michigan. WHK thanks the MCTP for
hospitality while part of this work was completed.


\begin{thebibliography}{99}

\bibitem{tolman} R. Tolman, {\it Relativity, Thermodynamics and
Cosmology} (Oxford U. Press, Clarendon Press, 1934).

\bibitem{steinturok} P. Steinhardt and N. Turok,
{\it Phys.Rev.} {\bf D65} 126003 (2002); J. Khoury, 
P. Steinhardt, and N. Turok, {\it Phys.Rev.Lett.} {\bf 92} 
031302 (2004).

\bibitem{phantom}
S.~M.~Carroll, M.~Hoffman and M.~Trodden,
Phys.\ Rev.\ D {\bf 68}, 023509 (2003);
A.~Melchiorri, L.~Mersini, C.~J.~Odman and M.~Trodden,
Phys.\ Rev.\ D {\bf 68}, 043509 (2003).

\bibitem{dickepeebles} R. Dicke and P.J.E. Peebles, in {\it General
Relativity: An Einstein Centenary Survey}, ed. by S. Hawking and W. Israel
(Cambridge: Cambridge Univ. Press, 1979).

\bibitem{davies} P. Davies, {\it Annales Poincare Phys.Theor.} {\bf
     49} 297 (1988); 
     E.~Babichev, V.~Dokuchaev and Y.~Eroshenko, arXiv:gr-qc/0402089.

\bibitem{ckw}  R. Caldwell, M. Kamionkowski, N. Weinberg,
  {\it Phys.Rev.Lett.} {\bf 91} 071301 (2003).

\bibitem{caldwell} R. Caldwell,
{\it Phys.Lett.} {\bf B545} 23 (2002).

\bibitem{shtanov}
Y.~Shtanov and V.~Sahni,
Phys.\ Lett.\ B {\bf 557}, 1 (2003)

\bibitem{branebounce}
P.~Kanti and K.~Tamvakis,
Phys.\ Rev.\ D {\bf 68}, 024014 (2003);
S.~Foffa,
Phys.\ Rev.\ D {\bf 68}, 043511 (2003).

\bibitem{singh}
A.~Ashtekar, T.~Pawlowski and P.~Singh,
  Phys.\ Rev.\  D {\bf 74}, 084003 (2006)
  [arXiv:gr-qc/0607039].

\bibitem{RSI} L. Randall and R. Sundrum,
{\it Phys.Rev.Lett.} {\bf 83}, 3370 (1999).

\bibitem{singularity}
D.~Lyth, Phys. Lett. {\bf B526}, 173 (2002);
J.~Martin, P.~Peter, N.~Pinto Neto and D.~J.~Schwarz, Phys.\ Rev.\ D {\bf  65}, 123513 (2002);
C.~Gordon and N.~Turok, Phys.\ Rev.\ D {\bf 67}, 123508 (2003);
J.~Martin and P.~Peter, Phys.\ Rev.\ Lett.\  {\bf 92}, 061301 (2004);
T.~J.~Battefeld, S.~P.~Patil and R.~Brandenberger, arXiv:hep-th/0401010.

\bibitem{cf} D. Chung and K. Freese,
  {\it Phys.Rev.} {\bf D61}, 023511 (2000)
 
\bibitem{braneworld}
  P. Binetruy, C. Deffayet, and D. Langlois,
{\it Nucl.Phys.} {\bf B565}, 269 (2000);
E.E. Flanagan, S. Tye, and I. Wasserman,
{\it Phys. Rev.} {\bf D62}, 024011 (2000);
C. Csaki, M. Graesser, C. Kolda, and J. Terning,
{\it Phys. Lett.} {\bf B462}, 34 (1999);
J. Cline, C. Grojean, and G. Servant, {\it Phys. Rev. Lett.}
{\bf 83}, 4245 (1999);
L. Mersini, {\it Mod. Phys. Lett.} {\bf A16}, 1583 (2001).

\bibitem{gw} 
W. Goldberger and M. Wise, {\it Phys. Lett.} {\bf B475}, 275
(2000).

\bibitem{allen}
L.~E.~Allen and D.~Wands, arXiv:astro-ph/0404441.

\bibitem{duality}
M.~P.~Dabrowski, T.~Stachowiak and M.~Szydlowski, Phys.\ Rev.\ D {\bf 68}, 103519 (2003);
L.~P.~Chimento and R.~Lazkoz, Phys.\ Rev.\ Lett.\  {\bf 91}, 211301 (2003);
J.~E.~Lidsey, arXiv:gr-qc/0405055;
L.~P.~Chimento and R.~Lazkoz, arXiv:astro-ph/040551.

\bibitem{Cline:2003gs}
   J.~M.~Cline, S.~Jeon and G.~D.~Moore,
   Phys.\ Rev.\  D {\bf 70}, 043543 (2004)
   [arXiv:hep-ph/0311312].

\bibitem{parker}
L.~Parker and A.~Raval,
  Phys.\ Rev.\ Lett.\  {\bf 86}, 749 (2001).

\end{thebibliography}
\end{document}